\documentclass[twocolumn,floatfix,aps,prl,showpacs,superscriptaddress,preprintnumbers,amsmath,amssymb]{revtex4}

\usepackage{graphicx}
\usepackage{bm}

\newcommand {\sNN}     {\mbox{$\sqrt{s_{\rm NN}}$}}
\newcommand {\kT}      {\mbox{$k_{\rm T}$}}
\newcommand {\mkT}     {\mbox{$\langle k_{\rm T}\rangle$}}
\newcommand {\Np}      {\mbox{$N_{\rm part}$}}
\newcommand {\Npc}     {\mbox{$N_{\rm part}^{1/3}$}}
\newcommand {\mNp}     {\mbox{$\langle N_{\rm part}\rangle$}}
\newcommand {\Rs}      {\mbox{$R_{\rm side}$}}
\newcommand {\Ro}      {\mbox{$R_{\rm out}$}}
\newcommand {\Rl}      {\mbox{$R_{\rm long}$}}
\newcommand {\Ros}     {\mbox{$R_{\rm out}/R_{\rm side}$}}
\newcommand {\chis}    {\mbox{$\chi ^{2}$}}
\newcommand {\qs}      {\mbox{$q_{\rm side}$}}
\newcommand {\qo}      {\mbox{$q_{\rm out}$}}
\newcommand {\ql}      {\mbox{$q_{\rm long}$}}

\newcommand {\bfq}     {\mbox{$\mathbf{q}$}}

\newcommand {\full}    {\mbox{\it{full}}}

\begin{document}

\title{Bose-Einstein Correlations of Charged Pion Pairs \\
       in Au+Au Collisions at $\sNN=200$ GeV.} 


\newcommand{\abilene}{Abilene Christian University, Abilene, TX 79699, USA}
\newcommand{\acadsin}{Institute of Physics, Academia Sinica, Taipei 11529, Taiwan}
\newcommand{\banaras}{Department of Physics, Banaras Hindu University, Varanasi 221005, India}
\newcommand{\barc}{Bhabha Atomic Research Centre, Bombay 400 085, India}
\newcommand{\bnl}{Brookhaven National Laboratory, Upton, NY 11973-5000, USA}
\newcommand{\caucr}{University of California - Riverside, Riverside, CA 92521, USA}
\newcommand{\ciae}{China Institute of Atomic Energy (CIAE), Beijing, People's Republic of China}
\newcommand{\cns}{Center for Nuclear Study, Graduate School of Science, University of Tokyo, 7-3-1 Hongo, Bunkyo, Tokyo 113-0033, Japan}
\newcommand{\columbia}{Columbia University, New York, NY 10027 and Nevis Laboratories, Irvington, NY 10533, USA}
\newcommand{\dapnia}{Dapnia, CEA Saclay, F-91191, Gif-sur-Yvette, France}
\newcommand{\debrecen}{Debrecen University, H-4010 Debrecen, Egyetem t{\'e}r 1, Hungary}
\newcommand{\fsu}{Florida State University, Tallahassee, FL 32306, USA}
\newcommand{\gsu}{Georgia State University, Atlanta, GA 30303, USA}
\newcommand{\hiroshima}{Hiroshima University, Kagamiyama, Higashi-Hiroshima 739-8526, Japan}
\newcommand{\ihepprot}{Institute for High Energy Physics (IHEP), Protvino, Russia}
\newcommand{\isu}{Iowa State University, Ames, IA 50011, USA}
\newcommand{\jinrdubna}{Joint Institute for Nuclear Research, 141980 Dubna, Moscow Region, Russia}
\newcommand{\kaeri}{KAERI, Cyclotron Application Laboratory, Seoul, South Korea}
\newcommand{\kangnung}{Kangnung National University, Kangnung 210-702, South Korea}
\newcommand{\kek}{KEK, High Energy Accelerator Research Organization, Tsukuba-shi, Ibaraki-ken 305-0801, Japan}
\newcommand{\kfki}{KFKI Research Institute for Particle and Nuclear Physics (RMKI), H-1525 Budapest 114, POBox 49, Hungary}
\newcommand{\korea}{Korea University, Seoul, 136-701, Korea}
\newcommand{\kurchatov}{Russian Research Center ``Kurchatov Institute", Moscow, Russia}
\newcommand{\kyoto}{Kyoto University, Kyoto 606, Japan}
\newcommand{\labllr}{Laboratoire Leprince-Ringuet, Ecole Polytechnique, CNRS-IN2P3, Route de Saclay, F-91128, Palaiseau, France}
\newcommand{\lawllnl}{Lawrence Livermore National Laboratory, Livermore, CA 94550, USA}
\newcommand{\losalamos}{Los Alamos National Laboratory, Los Alamos, NM 87545, USA}
\newcommand{\lpc}{LPC, Universit{\'e} Blaise Pascal, CNRS-IN2P3, Clermont-Fd, 63177 Aubiere Cedex, France}
\newcommand{\lund}{Department of Physics, Lund University, Box 118, SE-221 00 Lund, Sweden}
\newcommand{\muenster}{Institut f\"ur Kernphysik, University of Muenster, D-48149 Muenster, Germany}
\newcommand{\myongji}{Myongji University, Yongin, Kyonggido 449-728, Korea}
\newcommand{\nagasaki}{Nagasaki Institute of Applied Science, Nagasaki-shi, Nagasaki 851-0193, Japan}
\newcommand{\newmex}{University of New Mexico, Albuquerque, NM, USA}
\newcommand{\nmsu}{New Mexico State University, Las Cruces, NM 88003, USA}
\newcommand{\ornl}{Oak Ridge National Laboratory, Oak Ridge, TN 37831, USA}
\newcommand{\orsay}{IPN-Orsay, Universite Paris Sud, CNRS-IN2P3, BP1, F-91406, Orsay, France}
\newcommand{\pnpi}{PNPI, Petersburg Nuclear Physics Institute, Gatchina, Russia}
\newcommand{\riken}{RIKEN (The Institute of Physical and Chemical Research), Wako, Saitama 351-0198, JAPAN}
\newcommand{\rkrbrc}{RIKEN BNL Research Center, Brookhaven National Laboratory, Upton, NY 11973-5000, USA}
\newcommand{\saispbstu}{St. Petersburg State Technical University, St. Petersburg, Russia}
\newcommand{\saopaulo}{Universidade de S{\~a}o Paulo, Instituto de F\'{\i}sica, Caixa Postal 66318, S{\~a}o Paulo CEP05315-970, Brazil}
\newcommand{\seoulnat}{System Electronics Laboratory, Seoul National University, Seoul, South Korea}
\newcommand{\stonybrkc}{Chemistry Department, Stony Brook University, SUNY, Stony Brook, NY 11794-3400, USA}
\newcommand{\stonycrkp}{Department of Physics and Astronomy, Stony Brook University, SUNY, Stony Brook, NY 11794, USA}
\newcommand{\subatech}{SUBATECH (Ecole des Mines de Nantes, CNRS-IN2P3, Universit{\'e} de Nantes) BP 20722 - 44307, Nantes, France}
\newcommand{\tenn}{University of Tennessee, Knoxville, TN 37996, USA}
\newcommand{\titech}{Department of Physics, Tokyo Institute of Technology, Tokyo, 152-8551, Japan}
\newcommand{\tsukuba}{Institute of Physics, University of Tsukuba, Tsukuba, Ibaraki 305, Japan}
\newcommand{\vandy}{Vanderbilt University, Nashville, TN 37235, USA}
\newcommand{\waseda}{Waseda University, Advanced Research Institute for Science and Engineering, 17 Kikui-cho, Shinjuku-ku, Tokyo 162-0044, Japan}
\newcommand{\weizmann}{Weizmann Institute, Rehovot 76100, Israel}
\newcommand{\yonsei}{Yonsei University, IPAP, Seoul 120-749, Korea}
\affiliation{\abilene}
\affiliation{\acadsin}
\affiliation{\banaras}
\affiliation{\barc}
\affiliation{\bnl}
\affiliation{\caucr}
\affiliation{\ciae}
\affiliation{\cns}
\affiliation{\columbia}
\affiliation{\dapnia}
\affiliation{\debrecen}
\affiliation{\fsu}
\affiliation{\gsu}
\affiliation{\hiroshima}
\affiliation{\ihepprot}
\affiliation{\isu}
\affiliation{\jinrdubna}
\affiliation{\kaeri}
\affiliation{\kangnung}
\affiliation{\kek}
\affiliation{\kfki}
\affiliation{\korea}
\affiliation{\kurchatov}
\affiliation{\kyoto}
\affiliation{\labllr}
\affiliation{\lawllnl}
\affiliation{\losalamos}
\affiliation{\lpc}
\affiliation{\lund}
\affiliation{\muenster}
\affiliation{\myongji}
\affiliation{\nagasaki}
\affiliation{\newmex}
\affiliation{\nmsu}
\affiliation{\ornl}
\affiliation{\orsay}
\affiliation{\pnpi}
\affiliation{\riken}
\affiliation{\rkrbrc}
\affiliation{\saispbstu}
\affiliation{\saopaulo}
\affiliation{\seoulnat}
\affiliation{\stonybrkc}
\affiliation{\stonycrkp}
\affiliation{\subatech}
\affiliation{\tenn}
\affiliation{\titech}
\affiliation{\tsukuba}
\affiliation{\vandy}
\affiliation{\waseda}
\affiliation{\weizmann}
\affiliation{\yonsei}
\author{S.S.~Adler}	\affiliation{\bnl}
\author{S.~Afanasiev}	\affiliation{\jinrdubna}
\author{C.~Aidala}	\affiliation{\bnl}
\author{N.N.~Ajitanand}	\affiliation{\stonybrkc}
\author{Y.~Akiba}	\affiliation{\kek} \affiliation{\riken}
\author{J.~Alexander}	\affiliation{\stonybrkc}
\author{R.~Amirikas}	\affiliation{\fsu}
\author{L.~Aphecetche}	\affiliation{\subatech}
\author{S.H.~Aronson}	\affiliation{\bnl}
\author{R.~Averbeck}	\affiliation{\stonycrkp}
% ...
\collaboration{PHENIX Collaboration} \noaffiliation

\date{\today}

\begin{abstract}
Bose-Einstein correlations of identically charged pion pairs were
measured by the PHENIX experiment at mid-rapidity in Au+Au collisions
at \sNN~= 200 GeV.
The Bertsch-Pratt radius parameters were determined as a function
of the transverse momentum of the pair and as a function of the
centrality of the collision.
Using the \full~Coulomb correction, the ratio \Ros~is smaller than
unity for \mkT~from 0.25 to 1.2 GeV/c and for all measured
centralities.
However, using recently developed partial Coulomb correction methods,
we find that \Ros~is 0.8-1.1 for the measured \mkT~range, and
approximately constant at unity with the number of participants.
\end{abstract}

\maketitle

Following its origin in the study of proton-antiproton annihilations
\cite{GGLPP}, Bose-Einstein correlations have been extensively used to
measure source distributions in relativistic heavy ion collisions
\cite{WiedemannP,CsorgoP}.
These measurements were originally motivated by theoretical
predictions of a large source size and/or a long duration of particle
emission \cite{GyulassyP,BertschP2,PrattP2} -- which would result from a
softening of the equation-of-state in a first-order phase transition
to a quark-gluon plasma (QGP).
The technique of Bose-Einstein correlations is based upon quantum
statistical interference, but final state interactions such as Coulomb
repulsion modify the relative momentum distributions for pairs of
identical particles emanating from the collision region.
Both effects are included in multidimensional Gaussian fits to the
normalized relative momentum distributions yielding the fit parameters
which are the RMS-widths in each dimension, \Rl, \Rs,
\Ro~\cite{PrattP,BertschP}, also referred to as HBT radii.
A finite duration of emission leads to an effective elongation of one
of the HBT radii, \Ro, which is parallel to the mean transverse
momentum of pair, $\kT = ({\bf p}_{\rm 1T} + {\bf p}_{\rm 2T})/2$;
however, the other radii along the axes perpendicular to \kT~represent
the geometrical source sizes.
Therefore the \Ros~ratio could become larger than unity even for a
cylindrical particle source, if a long duration of particle emissions
occurs.
For dynamic (i.e. expanding) sources, the HBT radii depend on the \kT,
and correspond to lengths of homogeneity, regions of the source which
emit particles of similar momentum \cite{SinyukovP2}.
Measuring the \kT~dependence of the HBT radii provides essential
constraints on dynamical models that include the space-time evolution
of the source \cite{PrattP3,NA44P}.

Hydrodynamic models for the space-time evolution of a rehadronizing
QGP predicted that the measurement of the \Ros~ratio at moderate
values of \kT~provides a sensitive test of the long duration
of particle emission, a signal of a slowly burning first order phase
transition from QGP to hadrons at RHIC \cite{GyulassyP}.
Their predictions that the \Ros~ratio should reach $\sim$1.5
at \kT~of $\sim$0.5 GeV/c were not borne out by initial measurements
of Bose-Einstein correlations of pions from Au+Au collisions at \sNN~=
130 GeV \cite{PHENIX130P,STAR130P}.
This disagreement between theory and data has been called the ``RHIC
HBT puzzle'' \cite{GyulassyP2}.

We present here data on Bose-Einstein correlations of
charged pion pairs measured by the PHENIX experiment at RHIC for Au+Au
collisions at \sNN~= 200 GeV.
In this analysis, we adopt a recent fitting technique that provides
for a self-consistent treatment of the Coulomb final state interaction
for a source that is made up of a smaller core and a more extended
halo of long-lived resonances.
We introduce a new parameterization in which the strength of the
Coulomb interaction is constrained by the measured unlike-signed pion
correlation.

The PHENIX detector provides particle identification (PID)
capabilities for hadrons, leptons and photons over a wide momentum
range.
The setup of the PHENIX detector has been described in detail
elsewhere \cite{PHENIX-NIM}.
In this analysis, we use the west arm of the central spectrometer,
which covers the pseudorapidity region $|\eta|<0.35$ and
$\Delta\phi=\pi/2$ in azimuthal angle over a region of 0.2 GeV/c
$<\kT<$ 2.0 GeV/c.
The drift chamber (DC), at a radial distance between 2.0 m and
2.4 m, provides trajectory information in the azimuthal direction.
A pad chamber (PC1) at 2.5 m provides z-coordinate information.
Combining the DC and PC1 information, a track model provides
a 3-dimensional trajectory and momentum vector for charged particles.
The momentum resolution is $\delta p/p \simeq 0.7\% \oplus 1.0\%\times
p$ (GeV/c), where the first term is due to the multiple scattering
before the DC and the second term is the angular resolution of the DC.
For this analysis, the electromagnetic calorimeter (EMCal) provides the
time of arrival of particles at its front face located 5.1 m from the
beam axis. 
The timing resolution is approximately 400 psec for hadrons.
This analysis is based on a sample of ~34 million minimum bias events
taken with a magnetic field of 0.78 T$\cdot$m and triggered by the
coincidence of the Beam-Beam Counters (BBC) and Zero-Degree
Calorimeters (ZDC) -- corresponding to $92 \pm 2\%$ of the total
inelastic cross section of 6.8 b.
Event centrality is determined from the correlation between the BBC
multiplicity and the analog response of the ZDC.
About 23 million events are selected with a requirement that the
collision vertex measured by the BBC has $|z|<30$ cm.
Each track is required to have an associated hit on the EMCal within
2$\sigma$ of the track's projection to the EMCal, where $\sigma$ refers
to the resolution of the projection.
Charged particles are identified by the time-of-flight technique using
timing information between the BBC and the EMCal, combined with
momentum and flight length calculated by the track model.
Charged particles in the PID zone within 1.5 $\sigma$ of the ideal
squared-mass peak of pions but 1.5 $\sigma$ away from the kaon
bands are identified as pions.
After the track quality and PID cuts, $\sim$45 million positive pions
and $\sim$51 million negative pions are selected in a momentum range
from 0.2 to 2.0 GeV/c.

The pion correlation function is experimentally defined as
$C_{2}(\bfq)=A(\bfq)/B(\bfq)$, where $A(\bfq)$ is the measured two-pion
(actual pair) distribution of pair momentum difference \bfq, and
$B(\bfq)$ is the background pair (mixed pair) distribution generated
using mixed events from the same data sample. 
Event mixing is done selecting events that have similar multiplicities
and event vertices.
As in our previous analyses at \sNN~= 130 GeV \cite{PHENIX130P},
actual pairs within 1 cm in the beam direction ($\Delta Z_{DC}$) and
0.06 radians in azimuthal angle ($\Delta\phi_{DC}$) in the DC are
eliminated from the pair sample to remove ghost tracks, then within 5
cm in $\Delta Z_{DC}$ and 0.03 radians in $\Delta\phi_{DC}$ are also
eliminated to avoid a strongly inefficient region for pairs.
Since pairs which are close to one another in the EMCal are either
reconstructed as single hit clusters or are affected by
cluster-sharing, pairs whose hits are within 8 cm at the EMCal are
eliminated.
The background mixed pairs are subject to the same cuts as actual
pairs.
After the pair cuts, $\sim$110 million positive pion pairs and $\sim$140
million negative pion pairs remain.
The number of pairs is 40 times larger than the data sample used for
the PHENIX data at \sNN~= 130 GeV \cite{PHENIX130P}.
The systematic errors on the HBT radius parameters from the pair
cuts are evaluated by varying the pair cuts to be $\sim$2$\%$ for
\Rs~and \Rl, and $\sim$7$\%$ for \Ro.
The correction for the multi-track reconstruction efficiencies in the
DC and EMCal are determined by a GEANT-based \cite{GEANTP} Monte Carlo
(MC) simulation of the detector.
The pair efficiencies of the DC and EMCal are estimated by the ratio of
the distribution of separations of actual pairs to that of normalized
mixed pairs. Also the multiplicity dependent pair efficiencies are
estimated by an embedding technique: simulated pion pairs are
embedded in real events and the reconstruction efficiency for these
simulated pion pairs versus multiplicity is calculated.
The pair reconstruction inefficiencies in the DC and EMCal are corrected
by using pair efficiency factors estimated by the embedded MC
simulation.
Finally, we removed pairs within 0.005 radians in $\Delta\phi_{DC}$
where a pair inefficiency still remains even after the correction by
MC.
Correction for the residual HBT effect \cite{ZajcP} is estimated as a
systematic error.
The acceptance for this analysis is large, as a consequence the
systematic error is $\sim$1$\%$ on each radius parameter.

In order to compare directly to previous analyses, we apply the
standard ``\full'' strength Coulomb correction calculated from the
pair Coulomb wave function \cite{PrattP2} for a 3-D Gaussian
parameterization of the source using the radii obtained from the
previous fit.
We fit the Bose-Einstein correlation with the \full~Coulomb
correction to the the 1-D $q_{inv}$ parameterization,
$C_2(q_{\rm inv})=1+\lambda_{\rm inv}\exp(-R_{\rm inv}^2 q_{\rm
inv}^2)$, and the 3-D Bertsch-Pratt parameterization is given by
\begin{equation}
C_{2} =1+\lambda\exp(-R_{\rm side}^{2}q_{\rm side}^{2}-R_{\rm
out}^{2}q_{\rm out}^{2}-R_{\rm long}^{2}q_{\rm long}^{2})
\label{equ:BPEQ}.
\end{equation}
The relative momentum \bfq~is decomposed into $(\qs,\qo,\ql)$, where
the longitudinal component (\ql) is parallel to the beam axis, the
out component (\qo) is parallel to the mean transverse momentum of the
pair, ${\bf k}_T$ and the side component (\qs) is perpendicular to
both \ql~and \qo \cite{PrattP,BertschP}.
This analysis is performed in the Longitudinal Center-of-Mass System
(LCMS), where the mean longitudinal momentum of the pair vanishes. In
this frame, the duration of particle emission couples exclusively to
\qo.
In a general case, cross-terms may appear in Eq. \ref{equ:BPEQ}, but
they vanish in our measurement of central collisions at mid-rapidity
due to symmetry reasons \cite{ChapmanP}.

In the realistic source, however, many of relatively long-lived
particles  (e.g. $\eta$, $\eta^{\prime}$) which decay into pions and
have a Bose-Einstein interference too narrow to be resolved by
experiment also have a Coulomb interaction that is negligible.
For this reason, a new functional form was proposed \cite{SinyukovP}
which is based upon a Core-Halo picture of the source \cite{TamasP}
and assumes that the fraction of pairs, $\lambda$, which have
Bose-Einstein interference are the only pairs that contribute to the
Coulomb interaction.
\begin{equation}
C_{2}=C_{core}+C_{halo}=[\lambda (1+G)F]+[1-\lambda]
\label{equ:CHEQ}
\end{equation}
where $G$ corresponds to the Gaussian term in Eq. \ref{equ:BPEQ}, and
$F$ is the Coulomb correction term.
This correction method was recently adopted by CERES \cite{CERESP} and
STAR \cite{STARP}.  However, our formula differs slightly, because 
the momentum resolution correction to $\lambda$ (which is small 
according to MC simulation) is included in our systematic error,
rather than being incorporated into $F$.

\begin{figure}[b]
\includegraphics[width=1.0\linewidth]{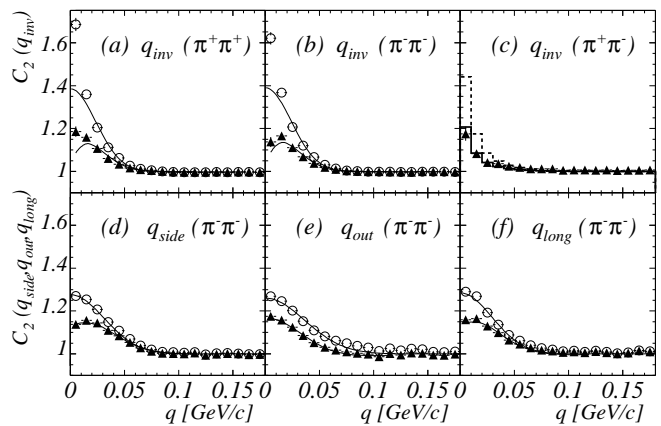}
\caption{\label{f:3DC2} 
Panels (a) and (b) show one-dimensional correlation
functions for $\pi^{+}\pi^{+}$ and $\pi^{-}\pi^{-}$.
The bottom figures show the three-dimensional correlation function for
$\pi^{-}\pi^{-}$ with the \full~Coulomb (opened circle) and without
Coulomb (filled triangle) corrections for 0.2 GeV/c $<\kT<$ 2.0 GeV/c
for 0-30$\%$ centrality.
The projection of the 3-D correlation functions are averaged over the
lowest 40 MeV in the orthogonal directions.
The error bars are statistical only.
The lines overlaid on the open circles (filled triangles) correspond
to fits to Eq. \ref{equ:BPEQ} (Eq. \ref{equ:CHEQ}) over the entire
distribution.
Panel (c) shows the one-dimensional correlation function of
unlike-signed pions for $0.2<\kT<2.0$ GeV/c.
The two overlaid histograms show calculations for the \full~(dashed)
and the $50\%$ partial (solid) Coulomb corrections.
}
\end{figure}

In order to test the underlying hypothesis of Eq. \ref{equ:CHEQ}, we
fit the strength of the Coulomb interaction ($\lambda_{+-}$) to the
unlike-signed correlation function in the range $0.2<\kT<2.0$ GeV/c,
and obtained a value for $\lambda_{+-}=0.50\pm0.04$ with
$\chis/DoF=3.0$.
This value is clearly inconsistent with the \full~strength Coulomb
correction but greater than the value of $\lambda$, especially for
$\kT<0.5$ GeV/c.
We attribute this difference mainly to the $\omega$ resonance, which
is sufficiently long-lived to be unresolved with Bose-Einstein
correlations, but may contribute significantly to the Coulomb
interaction.
To account for this contribution, we have modified Eq. \ref{equ:CHEQ}
to fix the total coulomb strength at $\lambda_{+-}$ while allowing the
Bose-Einstein fraction, $\lambda$, to vary from zero to
$\lambda_{+-}$,
\begin{equation}
C_{2}=[\lambda (1+G)F]+[(\lambda_{+-}-\lambda)F]+[1-\lambda_{+-}]
\label{equ:50EQ}
\end{equation}
In applying this formula, we still calculate the additional Coulomb
fraction using the Bertsch-Pratt source of approximately 5 fm, rather
than estimating the larger source distribution for the $\omega$ decay
products.
Therefore we use this formula to provide an upper bound on the effect
of the additional Coulomb interaction. The difference between
Eq. \ref{equ:50EQ} and Eq. \ref{equ:CHEQ} is used in our estimate of
the systematic errors.

Figure \ref{f:3DC2} shows the one-dimensional correlation function for
$\pi^{+}$ pairs, $\pi^{-}$ pairs, and unlike-signed pion pairs along
with projections of the three-dimensional correlation functions onto
\qs, \qo~and \ql~for $\pi^{-}$ pairs.

\begin{figure}[b]
\includegraphics[width=1.0\linewidth]{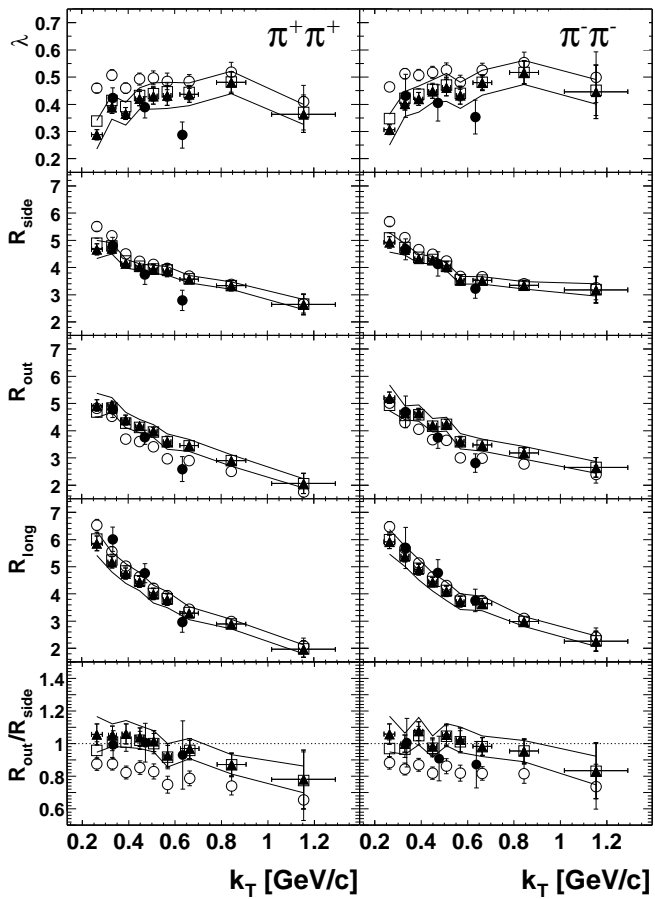}
\caption{\label{f:kTdep}
The \kT~dependence of the Bertsch-Pratt radius parameters
and $\lambda$ for charged pions for 0-30$\%$ centrality.
Filled triangles show the results from fits to a core-halo structure
by Eq. \ref{equ:CHEQ}, with statistical error bars and systematic
error bands.
Open circles and squares show the results from the
\full~(Eq. \ref{equ:BPEQ}) and 50$\%$ partial (Eq. \ref{equ:50EQ})
Coulomb corrections with statistical error bars, respectively.
Results at 130 GeV by PHENIX are given by filled circles.}
\end{figure}

Figure \ref{f:kTdep} shows the \kT-dependence of $\lambda$, Bertsch-Pratt
radii, and the ratio \Ros~for the $30\%$ most central events,
corresponding to \mNp~= 281.
For the \full~Coulomb correction, $\lambda$ is approximately constant
in all the \kT~bins while the radius parameters fall rapidly with
increasing \kT.
The \full~Coulomb corrected radius parameters at \sNN~= 200 GeV are
slightly different from the \full~Coulomb corrected radius
parameters at 130 GeV \cite{PHENIX130P} at the same \mkT because of
the improved pair efficiency correction, but similar within errors.
For the partial Coulomb correction, results from the two different
correction methods show similar trends. Compared to the \full~Coulomb
correction \Rs~and \Rl~systematically decrease while \Ro~increases,
and $\lambda$ decreases in the low \kT~region.
In the case of the \full~Coulomb correction, the ratio \Ros, in
Fig. \ref{f:kTdep}, is around 0.6-0.8 up to \kT$\sim$1.2 GeV/c.
On the other hand, \Ros~from the partial Coulomb correction is
systematically larger than that from the \full~Coulomb correction, and
slightly decreases from $\sim$1.1 to $\sim$0.8 as \kT~increases.
Reanalysis of the 130 GeV/c data with Eq. \ref{equ:CHEQ} gave results
that are fully consistent with the 200 GeV/c results.

\begin{figure}[b]
\includegraphics[width=1.0\linewidth]{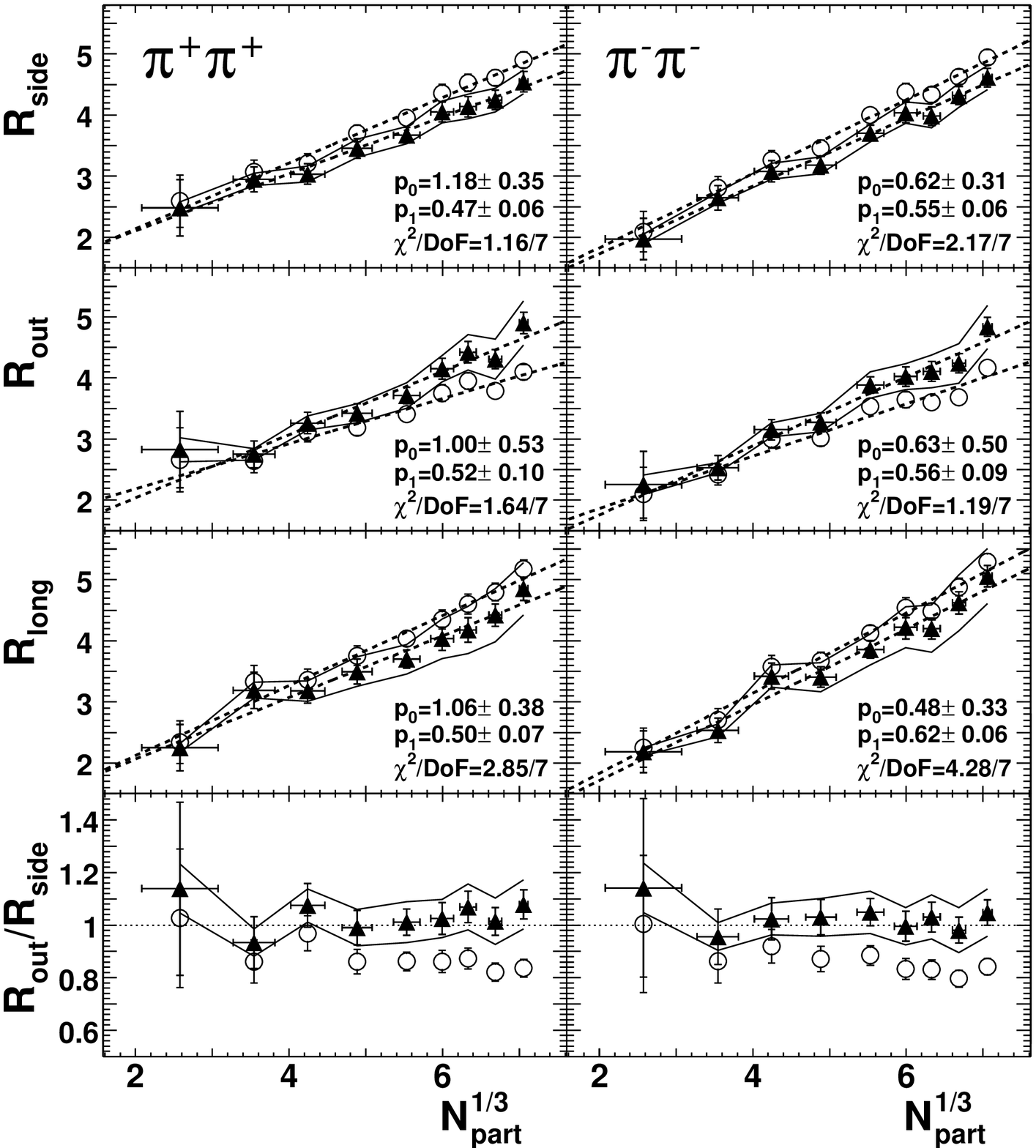}
\caption{Bertsch-Pratt radius parameters versus the cube root of the
number of participants (\Npc) for charged pions for the
fits to a core-halo structure by Eq. \ref{equ:CHEQ} (filled triangles)
with statistical error bars and systematic error bands, in the 0.2
GeV/c $<\kT<$ 2.0 GeV/c range, with \mkT~= 0.45 GeV/c. 
The dashed lines show fits to $p_{0}+p_{1}*\Npc$.
Fitted $p_{0}$ and $p_{1}$ values with fits to the core-halo stracture
are given in each panel.
The opened circles show results with the \full~Coulomb correction
by Eq. \ref{equ:BPEQ}.}
\label{f:Ctdep}
\end{figure}

Figure \ref{f:Ctdep} shows the collision centrality dependence of
the radius parameters. The number of participants (\Np) is
evaluated from the charged particle multiplicity using a Glauber
model calculation \cite{PHENIX130M}.
To evaluate the \Np~dependencies of the Bertsch-Pratt radius
parameters, we fit with a function of $p_{0}+p_{1}*\Npc$.
For the \full~Coulomb correction, the fit $p_{1}$ parameters
indicate that \Rs~and \Rl~show similar \Np~dependencies and \Ro~has a
slightly smaller \Np~dependence.
For the fits to a core-halo structure, all radius parameters show
similar \Np~dependencies.
All radii are consistent with a linear increase with \Npc.
The \Ros~ratios are approximately constant for all centralities.
The ratios from the partial Coulomb corrections are systematically
higher than those using the \full~Coulomb corrections.
These fit parameters are given in Fig. \ref{f:Ctdep}.

In conclusion, we have presented the Bertsch-Pratt HBT radii in the
LCMS for identified charged pions measured by PHENIX in Au+Au
collisions at \sNN~= 200 GeV.
The \kT~dependence of the HBT radii was measured for \mNp~= 281, and
the centrality dependence was measured for \kT~= 0.45 GeV/c.
These measurements are consistent with results from Au+Au collisions
at 130 GeV when a similar analysis (\full~Coulomb correction) is
performed.
We also performed two different partial Coulomb analyses, one based
upon a self-consistent treatment of the Coulomb correction, and the
other based upon direct comparison to the unlike-signed correlation,
which is shown to be inconsistent with the application of a
\full~Coulomb correction.
Both Coulomb corrections (Eq. \ref{equ:CHEQ} and Eq. \ref{equ:50EQ})
yield similar values of \Ros~which slightly decreases from $\sim$1.1
to $\sim$0.8 in the range of $\kT=0.2-1.2$ GeV/c for \mNp~= 281, and
approximately constant at unity with the number of participants for
\mkT~= 0.45 GeV/c.
These detailed measurements of the transverse momentum dependence of
the HBT radii, in particular that of \Ros, provide extremely
strong constraints for model builders.

We thank the staff of the Collider-Accelerator and Physics
Departments at BNL for their vital contributions.  We acknowledge
support from the Department of Energy and NSF (U.S.A.), MEXT and
JSPS (Japan), CNPq and FAPESP (Brazil), NSFC (China), CNRS-IN2P3
and CEA (France), BMBF, DAAD, and AvH (Germany), OTKA (Hungary),
DAE and DST (India), ISF (Israel), KRF and CHEP (Korea),
RMIST, RAS, and RMAE, (Russia), VR and KAW (Sweden), U.S. CRDF
for the FSU, US-Hungarian NSF-OTKA-MTA, and US-Israel BSF.


\end{document}